\documentstyle[12pt]{article}

\begin{document}

\begin{titlepage}
July, 1992 \hfill                 BNL-\\
\begin{center}

{\LARGE\bf The Phase Diagram of the $N=2$ Kazakov-Migdal Model}

\vspace{1.5cm}
\large{
Andreas Gocksch and Yue Shen\\
Physics Department, Brookhaven National Laboratory\\
Upton, NY 11973, USA \\

}
\vspace{1.9cm}
\end{center}

\abstract{
We have determined the phase diagram of the simplest version of a
lattice model introduced in the recent work of Kazakov and Migdal.
If $m_0$ and $\lambda$ are the bare mass and self coupling of the
scalar field in the model respectively, we find a line of first order phase
transitions in the ($m_0, \lambda$) plane ending in a critical point
where $\lambda$ is nonzero. Kazakov and Migdal speculate that their model of
scalar field theory could induce QCD.
Our work indicates that for $N=2$ there is no continuum limit for
the Kazakov-Migdal model except at the critical end point.
Whether or not a nontrivial continuum limit exists
in the vicinity of the critical point
requires careful study of the renormalization group properties of the model.
}
\vfill
\end{titlepage}

Recently Kazakov and Migdal (KM) \cite{KM} came up with an intriguing idea:
consider the following model, defined on a d-dimensional hypercubic
lattice with action
\begin{equation}
S = N \sum_x \left[ V(\Phi(x))-tr \sum_{\mu = 1}^d \Phi(x)U_\mu(x)
\Phi(x+\mu)U^\dagger_\mu(x)\right]~,
\end{equation}
where $\Phi(x)$ is a traceless scalar field in the adjoint
representation of $SU(N)$, covariantly coupled to the gauge fields
$U_\mu(x)$ defined on the links of the lattice. Kazakov and Migdal
speculate that perhaps, when the parameters of the scalar potential $V$
are tuned appropriately, the model Eq. (1) will have a non-trivial continuum
limit. Moreover the resulting continuum theory might actually be equivalent
to QCD.

If the above were true this apparent reformulation of QCD should
not be considered as merely an interesting curiosity.
The potential of this speculation lies in the fact that Eq. (1) in many ways
is much simpler than conventional lattice QCD. As a matter of fact, based on
the work of KM and of Migdal \cite{migdal} it looks as though in the limit
of large $N$ the model might be analytically tractable and that the degrees
of freedom associated with the eigenvalues of
$\Phi$ could actually be the long sought ``master field" of large $N$
QCD.

In order to see how such a scenario could possibly be true and also to
motivate the present study, let us examine Eq. (1) in more detail.
Integrating out the scalar fields in the partition function will give
rise to an induced action for the gauge fields which because of the
gauge invariance of Eq. (1), will be given as an (infinite) sum over Wilson
loops on the lattice. Moreover, because the scalar field is in the
adjoint representation, the Wilson loops will also be. Concretely, for
$V(\phi) = \frac {m_0^2}{2}tr{\phi}^2$, the $\phi$ integral is Gaussian
and one obtains the induced action\cite{KM}:
\begin{equation}
S_{ind} = - \sum_{\Gamma} \frac {2^{l(\Gamma)-1}|trU(\Gamma)|^2}
{l[\Gamma]m_0^{2l(\Gamma)}}~,
\end{equation}
where $l(\Gamma)$ is the length of the path $\Gamma$. Note that Eq. (2)
inherits
from Eq. (1) a {\it local} center symmetry $U_\mu(x) \rightarrow
Z_\mu(x)U_\mu(x)$, a fact recently discussed by Kogan, Semenoff and
Weiss \cite{kogan}. We will return to this point again later on.

How could the theory defined by Eq. (2) possibly be related to QCD? To
begin with the induced action in Eq. (2) is as good a lattice action as any
other in the sense that in the naive continuum limit, it produces a term
proportional to $F_{\mu\nu}^2$. If a continuum limit indeed exists for this
model, we expect that only this term will survive based on the familiar
renormalization group argument. The model is unusual only because the basic
building blocks of
the action are arbitrary loops ( in the adjoint representation).
In the weak smooth gauge field limit this model has a critical point at
$m_0^2 = 2d$ as can be seen easily from Eq. (1) where the lowest order
action becomes an action for a free scalar field. Taking fluctuations into
account the critical point is expected to shift to a different value $m_c^2$.
Using the definition $m^2 = m_0^2 - m_c^2$ KM \cite{KM} calculated the
coefficient of the $F_{\mu\nu}^2$ term and obtained the correspondence
\begin{equation}
\frac{1}{g_0^2} \rightarrow -{N\over 96\pi^2} ln(m^2a^2) ~,
\label{eq:bareg}
\end{equation}
in the smooth gauge fields  limit.

KM \cite{KM} and Migdal \cite{migdal} pointed out that the model in Eq. (1)
can be easier dealt with analytically and numerically if one
{\it first} integrates out the gauge fields. In the large $N$
limit, the resulting effective action for the master field is best
rewritten as a two-matrix model, which can be treated using modern
matrix model technology. KM suggest that the master field  gives
nontrivial scaling laws for physical quantities (for example, glueball
mass $M_g$)
\begin{equation}
M_g^2 =  (m^2)^{\gamma}~,
\end{equation}
where the value of $\gamma$ is different from the usual trivial scalar
theory ($\gamma = 1$ with possible logarithmic corrections).
KM have given the following argument for the value
of $\gamma$:
if the continuum theory of Eq. (1) is to ``induce" QCD correctly, we must
view the scalar as a {\it heavy} ``constituent" field whose mass
acts as an effective UV cutoff for the gauge field in the continuum.
This immediately gives the usual relation between a physical quantity
, the cutoff and the bare coupling
\begin{equation}
ln \frac{m^2}{M_g^2} = {48\pi^2 \over 11N g_0^2}~.
\end{equation}
Combining this expression with that of Eq. (\ref{eq:bareg}) one gets
the scaling relation for $M_g$ and find $\gamma = 23/22$ (This is the value
in the smooth gauge field limit. When the gauge field is treated
nonperturbatively $\gamma$ value can be quite larger than one \cite{migdal}).

Based on the above discussions we expect that
in order for Eq. (1) to correctly induce QCD, we need two things,
a critical point and nontrivial scaling of physical quantities.
In the present letter we follow KM's
suggestion and study the simplest version of the model, $SU(2)$ ($N=1$
is trivial). Certainly the possibility of Eq. (1) inducing QCD does not
crucially depend on $N$ being large. Integrating out the gauge fields
one obtains the following partition function \cite{KM}
\begin{equation}
Z = \int_{\phi >0}[d\phi]
\exp \left\{\sum_x \left[ ln{\phi}^2-2V(\phi)\right]+\sum_{<xy>}
ln\left[{sinh(4\phi(x)\phi(y))\over \phi(x)\phi(y)}\right]\right \}~,
\label{eq:scalar}
\end{equation}
where $<xy>$ denotes the sum over nearest neighbors of x.
The field $\phi(x)$ in Eq. (\ref{eq:scalar}) is the
(positive) eigenvalue of the matrix $\Phi$ and the restriction to $\phi(x)
>0$ expresses the fact that it is a ``radial" variable. According to what
was said before, we must find a critical point close to which a continuum
limit can be constructed. It is well known that the continuum limit of
a scalar model with polynomial interaction terms is trivial with vanishing
renormalized coupling \cite{wilson}.
In principle the continuum limit of
Eq. (\ref{eq:scalar}) could avoid the triviality problem
due to the presence of the nonpolynomial interaction terms (if we
could integrate out the angular variables in an $O(N)$ model exactly, however,
 the resulting effective action for the radial variables would be also
nonpolynomial).
Whether or not this is the case and Eq. (\ref{eq:scalar}) allows for
nontrivial scaling is numerically a much more difficult
question and remains to be answered in the future.

We restrict ourselves to a scalar potential of the form $V(\phi) = \frac
{m_0^2}{2}tr\phi^2 + \frac{\lambda}{4}tr\phi^4$.
A nonzero $\lambda$ is necessary to stablize the system and allows numerical
simulation in the small $m_0^2$ region. In order
to get an idea of what the phase diagram looks like in the ($\lambda,
m_0$) plane, we first use a simple mean field theory (MFT), i.e. we set
$ \phi(x)= e^{u}$ and assume $u(x) = constant$. We then analyzed the resulting
potential looking for minima and phase transitions.
As it turns out the qualitative features of the phase diagram from the MFT
effective potential analysis are in
excellent agreement with Monte Carlo simulations.
In Fig. 1, we show a typical plot of the MFT
potential $V_{eff}(u)$ at a rather small value of $\lambda = 0.01$. There
obviously is a first order phase transition at $m_0^2 = m_c^2= 7.65$,
which is not very far away from the smooth gauge field limit $m_0^2 = 8$. For
this particular value of $\lambda$ we show a comparison of MFT and Monte
Carlo for $\phi$-field vacuum expectation value in Fig. 2.
All our Monte Carlo simulations were done on $8^4$
lattice using a single-hit Metropolis algorithm on the field $u$.
The errors in Fig. 2 were obtained by blocking averages and $<\phi>$ values
are the result of $5K$ sweeps through the lattice at each point.
Clearly, the agreement between MFT and Monte Carlo is very good.
The transition is obviously
first order with a very clear hysteresis in $<\phi>$ (we also have
obtained clear two state signals at $m_c^2$). Interestingly then,
our first conclusion is that in the $\lambda \to 0$ limit
{\it no} continuum limit exists.

The appearance of first order phase transitions does not come as a surprise.
The simple $SU(2)$ lattice theory in the adjoint representation
($SU(2)/Z_2 = SO(3)$) has action $S = \beta_A \sum_D |TrU_D|^2$ and
possesses a first order phase transition at some critical value of $\beta_A$
\cite{greensite}. Presumably this transition is related to the $Z_2$ invariance
of the action \cite{kogan}. The question now is whether a first order
transition
can be avoided by tuning $\lambda$. The answer is yes: In MFT the gap in
$u$ (or $<\phi>$) decreases rapidly as $\lambda$ is increased. For
$\lambda > 2.57$ the gap disappears all together and there is no phase
transition anymore. Thus the phases of strong and weak $m_0^2$ are analytically
connected. The point ($\lambda = 2.57, m_0^2 = 4.515$) is a
critical point in MFT: $V_{eff}^{\prime\prime}(\phi)$ vanishes at the minimum
of the potential. In Fig. 3 we show how this MFT scenario is confirmed by Monte
Carlo simulations. The location of the critical endpoint by Monte Carlo
is not very precise. It is taken at the value of $\lambda$ where the hysteresis
effect in $<\phi>$ disappears and the critical value of $m_0^2$ is
determined by the position of the peak of $\phi$-field susceptibility
\begin{equation}
\chi = {1\over V} \sum_{x,y}\left[ <\phi(x)\phi(y)> -
<\phi(x)><\phi(y)>\right]~.
\end{equation}

In conclusion, we have found a line of first order phase transitions in
the ($\lambda, m_0^2$) plane of the SU(2) KM model. There is no continuum
limit at $\lambda \to 0$. The first order phase transition line terminates in a
critical point at $\lambda > 0$, in the vicinity of which one may attempt
to construct a continuum theory. The precise nature of this continuum theory
remains to be determined. This is a challenge for further research.

\noindent{Acknowledgement}

We thank V. Emery and K. Jansen for valuable discussions.
This work was supported by DOE grant at
Brookhaven National Laboratory (DE-AC02-76CH00016). The numerical
simulations are performed at SCRI at Tallahassee.

\pagebreak

\section*{Figure Caption}

{\bf Figure 1}: The MFT effective potential $V_{eff}(u)$ at $\lambda=0.01$.
The two minima become degenerate at $m_{0}^2 = 7.65$ which agrees
with the observed critical point for the first order phase transition
in Monte Carlo simulation.

\noindent {\bf Figure 2}: Comparison of the mean field theory and
Monte Carlo simulation result for $\phi$-field vacuum expectation value.
The solid lines are the MFT predictions and
Monte Carlo data points are indicated by *. The errors are smaller than the
size of the symbols.

\noindent {\bf Figure 3}: The phase diagram. The solid line is the mean
field theory prediction and the simulation results are indicated by diamonds.
The end point of the first order phase transition line is marked by *
for MFT and square for Monte Carlo estimate. At the end point the theory
becomes critical.

\pagebreak


\begin{thebibliography}{99}

\bibitem{KM} V.A. Kazakov and A.A. Migdal, Princeton Univ. preprint
PUPT-1322 (1992).

\bibitem{migdal} A.A. Migdal, Princeton Univ. preprint PUPT-1323 (1992).

\bibitem{kogan} I.I. Kogan, V.W. Semenoff and N. Weiss, UBC preprint UBCTP
92-022.

\bibitem{wilson} K. G. Wilson and J. Kogut, Phys. Repts. 12, (1974) 75.

\bibitem{greensite} J. Greensite and B. Lautrup, Phys. Rev. Lett. 47, (1981) 9;
E.~G.~Halliday and A. Schwimmer, Phys. Lett. 101B, (1981) 327.

\end{thebibliography}
\end{document}